\documentclass[letterpaper,12pt]{article}   

\usepackage{graphics, epsfig}
\begin{document}

\title{Comment on `` Topological phase in two flavor neutrino oscillations"}


\author{\bf Rajendra Bhandari}
\maketitle
\vspace{10mm}
\begin{center}
\begin{tabular}{ll}
            & Raman Research Institute*, \\
            & Bangalore 560 080, India. \\
            & email: bhandari@rri.res.in,\\
            &       rajbhand@yahoo.com\\
\end{tabular}
\end{center}
\vspace{10mm}

\begin{center}
\Large{\bf {Abstract}}
\end{center}
We critically analyze the claims with regard to the relevance of topological phases in the physics of neutrino oscillation made  in a recent paper [Phys. Rev. D 79, 096013 (2009)]  
and point out some inappropriate exaggerations and misleading statements. We find that the $\pi$ phase described in this paper, while interesting, is an artefact of two major 
approximations made in the paper. We point out a more robust and more familiar $\pi$ phase in the neutrino oscillation formulae which 
can be interpreted as a pure Pancharatnam phase.  We also make some relevant remarks on the distinction between the geometric and the topological phase made in the commented paper.
\vspace{5mm}

\newpage
 

In a recent paper \cite{mehta} it has been claimed that `` for the minimal case of two flavors and CP conservation, there is a geometric interpretation 
of the neutrino oscillation formulae for the survival and detection probabilities of neutrino 
species". In this paper we first recall the derivation of the main result in \cite{mehta}with a slightly different notation
and then show that (i) there is a more robust and more familiar $\pi$ phase in neutron oscillation formulae than the one discussed in  \cite{mehta}  which can be seen 
as a pure Pancharatnam phase, (ii) the geometric interpretation and the $\pi$ phase discussed in  \cite{mehta} is an artefact of two important approximations made in the paper 
and thus lacks fundamental significance, (iii) the geometric interpretation belongs  only to an aspect of the neutron oscillation formulae and 
not to the formulae themselves,  (iv) the distinction between the topological 
phase and geometric phase introduced in the paper is inappropriate and (v) discuss certain misleading statements made in the paper.

As in \cite{mehta}, let $|\nu_{\alpha}>$ and $|\nu_{\beta}>$ be the two flavor eigenstates 
represented by the two antipodal points lying along the z-axis and let 
$|\theta,\pm>$ be the two orthogonal mass eigenstates which lie on the line making an angle $\theta$ 
with respect to the z-axis. Let  the initial state be

\begin{equation}
|\nu_{\alpha}>=c_+|\theta_1,+> + c_-|\theta_1,->,\label{eq:1}  
\end{equation}

\noindent where the coefficients $\nu_{\alpha +}$ and $\nu_{\alpha -}$ of \cite{mehta}
have been replaced by $c_+$ and $c_-$, the rest of the   being the same. This 
state evolves in time t to the state 

\begin{equation}
|\nu_{\alpha}>'=e^{iD_+}c_+|\theta_2,+> + e^{iD_-} c_-|\theta_2,->,\label{eq:2}  
\end{equation}

\noindent where $D_+$ and $D_-$ are the dynamical phases acquired during the 
adiabatic evolution of the states $|\theta, \pm>$ from $|\theta_1,\pm>$ to 
$|\theta_2,\pm>$.

The survival probability $P_\alpha$ and the transition probability $P_\beta$ are given by, 

\begin{eqnarray}
P_{\alpha}={|<\nu_{\alpha}|\nu_{\alpha}>'|}^2=|c_+|^2|<\nu_{\alpha}|\theta_2,+>|^2 + |c_-|^2|<\nu_{\alpha}|\theta_2,->|^2 \nonumber \\
 +({c_+}^* c_- e^{i(-D_- +D_+)} {<\nu_{\alpha}|\theta_2,+>}^* <\nu_{\alpha}|\theta_2,-> + c.c.), \label{eq:3} 
\end{eqnarray}

\begin{eqnarray}
P_{\beta}={|<\nu_{\beta}|\nu_{\alpha}>'|}^2=|c_+|^2|<\nu_{\beta}|\theta_2,+>|^2 + |c_-|^2|<\nu_{\beta}|\theta_2,->|^2 \nonumber \\
 +({c_+}^* c_- e^{i(-D_- +D_+)} {<\nu_{\beta}|\theta_2,+>}^* <\nu_{\beta}|\theta_2,-> + c.c.).\label{eq:4} 
\end{eqnarray}

We first note a rigorous result within the two-flavor model which does not depend on any approximations.
Since it has been assumed that there is no decay, $P_{\alpha}$ and $P_{\beta}$ must add to 1. An 
inspection of eqns. (3) and (4) makes it obvious that this is possible only if the cross terms 
in the two equations add to zero. This implies that the complex numbers 
$A_\alpha= {<\nu_{\alpha}|\theta_2,+>}^* <\nu_{\alpha}|\theta_2,->$ and 
$A_\beta= {<\nu_{\beta}|\theta_2,+>}^* <\nu_{\beta}|\theta_2,->$ should 
be equal in magnitude and differ in phase by $\pi$, i.e. the phase of the number ${A_\beta}^*{A_\alpha}$
must be equal to $\pm\pi$. Now 

 \begin{eqnarray}
{A_\beta}^*{A_\alpha}={<\nu_{\beta}|\theta_2,+>} {<\nu_{\beta}|\theta_2,->}^* {<\nu_{\alpha}|\theta_2,+>}^* <\nu_{\alpha}|\theta_2,->. \label{eq:5}
\end{eqnarray}
\noindent Using the fact that ${<\nu_{\beta}|\theta_2,->}^*=<\theta_2,-|\nu_{\beta}>$ etc. and 
rearranging the terms we get

 \begin{eqnarray}
{A_\beta}^*{A_\alpha}={<\nu_{\beta}|\theta_2,+>}<\theta_2,+|\nu_{\alpha}><\nu_{\alpha}|\theta_2,-> <\theta_2,-|\nu_{\beta}>. \label{eq:6}
\end{eqnarray}

\noindent By Pancharatnam's theorem, the phase of the complex number on the right hand 
side of Eq.(6) is equal to half the solid angle subtended by the closed geodesic curve 
starting at the state $|\nu_{\beta}>$, passing through thestates $|\theta_2,->$, $|\nu_{\alpha}>$, 
$|\theta_2,+>$ and ending at $|\nu_{\beta}>$, which is a great circle on the Poincar\'{e} sphere. This phase is obviously equal 
in magnitude to $\pi$. This is the well known $\pi$ phase between the oscillations of intensities of the two different 
flavours which is seen here as a pure Pancharatnam phase of magnitude $\pi$. It follows from the condition of 
unitarity  and
is an elegant example of internal consistency of different principles of physics 
having apparently different origins. Within the two-state model, this phase   is independent of any approximations. It 
can be shown easily that this $\pi$ phase does not depend even on the adiabatic approximation.
 Let us also note that (a) this phase is independent of the 
phases of the individual states occurring in Eq.(6) and (b) this phase can 
be looked upon as the phase acquired by the state $|\nu_{\alpha}>$ if it evolved along the closed great 
circle under the action of a constant unitary hamiltonian that represents rotation about an axis 
perpendicular to the great circle, i.e. under an SU(2) element that represents a $2\pi$ rotation on the 
 Poincar\'{e} sphere about this axis. 

If one stares at Eqs.(3) and (4) for a while it becomes obvious that the content of the above 
result can be exactly simulated by the following polarization experiment: Let polarization states 
$|\theta_2,+>$ and $|\theta_2,->$ be incident on the two slits of an interferometer and a polarizer that 
passes the state $|\nu_{\alpha}>$ be placed in front of the screen and the position of the fringes 
noted. Let the polarizer  now be replaced by one that passes the state $|\nu_{\beta}>$ and the shift 
in the fringes measured. The above result says that irrespective of the incident states  the measured 
phase shift must be equal to $\pi$ in magnitude.

The above result has in fact been demonstrated in optical interference experiments \cite{iwbs}
using the two-state system of light polarization. The 
results shown in Fig. 3 of \cite{iwbs} show that a rotation of a linear polarizer 
through $90^\circ$ always results in a phase shift of $\pm \pi$ irrespective of the 
polarization states of the interfering beams.

We next come to the $\pi$ phase discussed by Mehta  \cite{mehta}.  In the cross term on the right hand side of 
Eq. (4), if we substitute for $c_+$ and  $c_-$ from Eq.(1) we get, after rearranging the terms, the product 
${<\nu_{\alpha}|\theta_1,+>}<\theta_2,+|\nu_{\beta}><\nu_{\beta}|\theta_2,-> <\theta_1,-|\nu_{\alpha}>$
multiplying the exponential term. To make this product correspond to 
evolution of a state along a closed great circle one needs two more terms $<\theta_2,-|\theta_1,->$  and 
$<\theta_1,+|\theta_2,+>$ which are missing from the product. To compensate for the missing terms, the 
author first sacrifices the arbitrariness of the phases of the individual states in the product and then  makes the 
approximation that the hamiltonian is CP non-violating. Let us note that the approximation of adiabatic evoution of the states 
 $|\theta_1,->$ and  $|\theta_1,+>$ to the states  $|\theta_2,->$ and  $|\theta_2,+>$ has already been made.  Under these 
two approximations, the author argues rightly that the phases of the missing terms are accounted for exactly if 
the phases of the pairs of states $|\theta_1,->$,  $|\theta_2,->$ and $|\theta_1,+>$,  $|\theta_2,+>$ in the product
are related by parallel transport and that the phase of the product  is then equivalent to that acquired in a unitary evolution along a great 
circle under a constant hamiltonian, i.e. equal in magnitude to $\pi$. If any of the two approximations is dropped the result is no 
more true. In fact the author shows in a separate paper \cite{mehta2} that if the adiabatic approximation is 
retained but the hamiltonian is allowed to be CP-violating, the product is no longer equivalent to evolution along a 
closed great circle and the  phase of the product is no longer $\pi$ but is equal to that determined by the 
solid angle of the distorted curve! The phase of magnitude $\pi$ is thus an artefact of the restriction on 
the hamiltonian. 

In order to evaluate the significance of the result let us first consider the case of evolution in free space or 
in constant density matter where the states $|\theta_1>$   and  $|\theta_2>$ are the same. In this case the 
variations of the flavor intensities are pure sinusoidal oscillations. The sinusoidal oscillation has three attributes:
(A) amplitue of the oscillation which is determined by the modulus of the cross terms in Eqs. (3) and (4), 
(B) frequency of the oscillation which is determined by the dynamical phase term containing $D_+$ and $D_-$ and (C) a constant phase which is 
equal to $0$ in case of the survival probability given by Eq. (3) and equal in magnitude to $\pi$ in case of the transition probability given by Eq. (4). 
The main results of the  paper are concerned with only the
attribute (C), i.e. the constant phase of the oscillation.  Someone who had never heard of 
the Pancharatnam phase would fix this constant phase trivially by requiring that the survival probability 
$P_{\alpha}$ and the transition probability $P_{\beta}$ be equal to 1 and 0 respectively at time t=0. 
Surely it can be seen as a Pancharatnam phase,  
but the claim on this basis, as in the abstract of the paper,  that ``the neutron oscillation formulae have 
a geometric interpretation" is, in our view, a gross exaggeration.
The statement ``More precisely, the standard result for neutrino oscillations is in fact 
a realization of the Pancharatnam topological phase"  on page 9 is also an 
inappropriate exaggeration. It is equivalent to claiming in the 
first example discussed in this comment that unitarity is a realization of the Pancharatnam phase !.
In our view the frequency of the oscillation determined by the dynamical phase and the amplitude of the 
oscillation determined by the mixing angle are at least  as important parts of the neutron oscillation formulae as 
the absolute phase of the oscillation.

 When variable matter density is introduced, the dynamical phase is no more a linearly varying function of time nor 
is the amplitude of the cross term constant in time. Both these effects are just as important to the variation of 
flavour intensities along the path as the geometric part of the phase and are parts of the neutron oscillation formulae. 
In fact in the adiabatic limit, since the dynamical phase is large compared to the geometric phase by definition, the variation 
of dynamical phase due to the presence of variable matter density could easily dominate over the geometric term.
Moreover, as discussed above, the value $\pi$ for the phase 
found by the author is  a consequence of the restriction on the hamiltonian and the adiabatic approximation and is , therefore, 
 not fundamental. It is just a special value obtained under  
specified constraints.

 A distinction  in 
terminology between the $\pi$ phase obtained when CP violation is absent and the 
non-$\pi$ phase obtained when it is present has been made in the paper, the 
former being called topological and the latter geometric. We find this distinction unnatural and inappropriate 
since both are parts of a single phenomenon and are manifestations of the same basic singularity associated with the 
SU(2) group. A unified description of this singularity has been described in several papers. In the context of 
adiabatic evolution, a three-dimensional generalization of the sign change rule has been 
described in \cite{rbmonopole}. In the context of nonadiabatic evolution an operational description of this 
singularity has been given in \cite{jumps,4pism,nonmodph09} and experimental demonstrations with polarization of light 
have been reported in \cite{4pism,iwbs,dirac1}. We  briefly recall this work below.

Consider the evolution of a spin-1/2 state under the action of a hamiltonian which is a function of three 
parameters $x$, $y$ and $z$. First let us consider evolution along a closed circuit  in a two-dimensional space, i.e. the plane $y=0$.
Let the hamiltonian be degenerate at an isolated point ($x_0,0,z_0$)  in this plane. Let the parameters of the 
hamiltonian be changed so that the closed circuit moves from a condition where it does not encircle the point 
 ($x_0,0,z_0$) to a condition where it does. Then the considerations in \cite{mehta} say that at the transition 
point where the boundary of the closed circuit crosses the degeneracy, there  is a sudden jump  of magnitude $\pi$ in the phase 
of the state, assuming that the dynamical phase has been subtracted out. We point out that at the point of crossing of the 
degeneracy the adiabatic approximation must break down.
It was pointed out in \cite{rbmonopole} that if  the same circuit  were located in the plane $y=\epsilon$, where $\epsilon$ is small and the same motion 
of the circuit carried out so that this time there is no actual crossing of the degeneracy, there is a measurable phase shift of a magnitude 
approximately equal to,  $+\pi$  whereas if  the same operation were carried out in the plane $y=-\epsilon$ there is a 
 measurable phase shift of a magnitude approximately equal to  $-\pi$ (only the relative sign being important) \cite{footnote}. It was further shown that if the circuit were taken around 
a closed loop such that it loops the degeneracy, there is a measurable phase shift equal to $\pm 2\pi$. A monopole of  strength 1/2  located at 
the degeneracy point gives a good unified description of the two effects which have been termed differently in \cite{mehta}.

To consider the more general nonadiabatic case, an element of SU(2) which corresponds to a $2\pi$ rotation about any axis on the Poincar\'{e} sphere 
is represented by the matrix $-{\bf 1}$.  Any state acted upon by this element therefore  acquires a phase 
of magnitude $\pi$. This is the well-known phenomenon of ``$4\pi$ spinor symmetry". States with different polar 
angles with respect to the rotation axis execute small circles of different diameter but the total phase acquired in one full cycle 
is always of magnitude $\pi$. There is, however, another dimension to the problem. A phase shift has a magnitude as well as 
a sign. If the phase of a state evolving under the above hamiltonian were continuously monitored with 
an interferometer with reference to some reference state  $|R>$, the total measured phase shift will be 
$+\pi$ for some states and $-\pi$ for others.
At critical values of the parameters, the phase shift jumps suddenly from being $+\pi$ to being $-\pi$.  
This has been verified in polarization experiments \cite{4pism,iwbs}.
This is due to the presence of phase singularities which can be identified as follows. Let $|R>$ stand 
for the reference state in the interferometer and $|{\tilde R}>$ the diametrically opposite state on the 
sphere which is orthogonal to  $|R>$. At points during the evolution where the evolving state is 
equal to $|{\tilde R}>$ the interference pattern vanishes and the phase is undetermined. In the 
vicinity of this point the phase shifts vary sharply. In general, if a state  $|I>$ incident on one arm of 
an interferometer undergoes an SU(2) transformation U which is a function of some variable parameters to 
yield a final state  $|F>$ which interferes with a state  $|R>$ in the reference arm  then 
a closed cycle of the parameters of U around an isolated point at which  $|F>$= $|{\tilde R}>$ yields a total phase shift 
equal to $\pm 2n\pi$ where $n$ is an integer index representing the strength of the singularity. Some examples of such phase shifts have been demonstrated in interference experiments \cite{dirac1}. 
This result is  conceptually simpler than the adiabatic result as it does not depend on subtraction of a large dynamical phase from the total phase. 
Note that in both the  above discussions the phase shift associated with the singularity is of magnitude $2n\pi$ and not $\pi$.
.

Finally we point out that at the end of page 2 of \cite{mehta}, the statement 
``Now this open loop (noncyclic) Schrodinger evolution of a quantum state
over a time $\tau$ can be closed by a collapse of the time-evolved quantum
state at $\tau$ onto the
original state at $\tau=0$ by the shortest geodesic curve joining the two
states in the ray space \cite{jsrb}."  is a misrepresentation of historical 
facts. A 
similar statement made by Samuel on page 960 of \cite{samuel} is also false. 
What was said in Samuel and Bhandari \cite{jsrb} 
was just the opposite.
It was stated repeatedly on page 2341 of \cite{jsrb} that the final state can be
connected to the initial state by {\it any} geodesic arc. After a careful discussion with the first author of  \cite{jsrb} the author of \cite{mehta} has stated in 
\cite{mehtareply} that the word ``any" in ``any geodesic arc" refers to any of the gauge copies of the 
geodesic in the $\it N$ - space. Our response to this is that the expression ``any geodesic arc" includes all geodesic arcs 
in $\it N$, i.e. those that project down to the shorter geodesic as well as those that project down to the 
longer geodesic in the ray space. The footnote from  \cite{jsrb}  cited in \cite{mehtareply} is merely a 
description of a property of the shorter geodesic and does not constitute a restriction on the definition of the noncyclic geometric phase. 
In fact the footnote  does not form part of the  discussion of the noncyclic geometric phase on p. 2341 where the expression 
``any geodesic curve" occurs again.
In response to the following comment made in \cite{mehtareply} : ``the author has been caught in the 
unfortunate position of having been scooped by himself", a statement we do not understand,  we reiterate that the first clear 
statement that the noncyclic geometric phase should be defined as half the solid angle of the area obtained  by closing the open curve in the ray space with the shortest geodesic arc connecting the final state to the initial state was made in 
 \cite{jumps}, i.e. [R. Bhandari,  Phys. Lett. {\bf A 157}, 221 (1991)] and not in  \cite{jsrb}, i.e. 
[J. Samuel and R. Bhandari,  Phys. Rev. Lett. {\bf 60}, 2339 (1988)]. This led to the prediction of observable $\pm \pi$ 
phase jumps in two-state systems which were later verified in interference experiments \cite{4pism,iwbs} and to the nonmodular view of the topological phase, thus 
constituting a conceptual advance in the subject. The restriction to the shorter geodesic is thus much more than a footnote. It may also be pointed 
out that the definition of  the noncyclic 
geometric phase proposed in \cite{jumps} as the difference of the total Pancharatnam phase of the evolving state and the dynamical phase 
as defined by Aharonov and Anandan does not depend on a geodesic rule and is thus particularly useful for 
systems with more than two states where the geometry of the ray space can not be easily visualized. An extension of the definition to the case of 
an arbitrary reference state has been proposed in \cite{nonmodph09}.
To end this discussion we  note that the fact that the shortest geodesic rule was not used in \cite{jsrb} was also pointed out in \cite{rbmonopole}.

To sum up, the net content of \cite{mehta} would be precisely summarized if the abstract of the paper read: 
``We show that, under the adiabatic approximation, the phase appearing in the neutrino oscillation formulae has a geometric 
contribution which, under the constraint of CP non-violation, is equal in magnitude to $\pi$ ". 
Considering that the phase in a quantum evolution in general  has   a geometric part, this is not very significant.

\vspace{5 mm}

\noindent {\bf \Large Acknowledgements}:\\

I thank the author of the criticized paper for clarifying her work in \cite{mehtareply}.



\begin{thebibliography}{4pism}
\bibitem{footnote}
The actual problem studied in \cite{rbmonopole} was a closely related one where the dynamical phase cancels due to the 
geometry of the experiment.
\end{thebibliography}
\end{document}